\documentstyle[12pt]{article}
\begin{document}

\title{An extension of Anderson-Higgs mechanism}
\author{Guo-Zhu Liu$^{1,3}\thanks{E-mail:gzliu@mail.ustc.edu.cn}$ and Geng Cheng$^{2,3}\thanks{E-mail:gcheng@ustc.edu.cn}$ \\
$^{1}${\small {\it Structure Research Laboratory, }}\\
{\small {\it University of Science and Technology of China, }}\\
{\small {\it Hefei, Anhui, 230026, P.R. China }}\\
$^{2}${\small {\it CCAST (World Laboratory), P.O. Box 8730, Beijing 100080, P.R. China }}\\
{\small {\it $^{3}$Department of Astronomy and Applied Physics, }}\\
{\small {\it University of Science and Technology of China, }}\\
{\small {\it Hefei, Anhui, 230026, P.R. China}}}
\maketitle

\begin{abstract}
\baselineskip20pt When free vortices are present in the pseudogap region of underdoped cuprates, it is shown that conventional Anderson-Higgs mechanism does not work because the Goldstone field is not an analytic function. However, after decomposing the Goldstone field into longitudinal and transverse components, we find that the former can be eliminated by a special gauge transformation and the gauge field becomes massive, but the latter persists in the theory in any case. Thus we obtain an extended Anderson-Higgs mechanism which exhibits completely new physics comparing with the conventional one.
\newline
\end{abstract}

\newpage

\baselineskip20pt

Spontaneous symmetry breaking, which occurs when an invariance of the Hamiltonian of a physical system is not an invariance of its ground state, is one of the most important concepts in modern physics [1]. According to the well-known Goldstone theorem [2], for every spontaneously broken global continuous symmetry there must be a massless particle called Goldstone boson. If we include both local gauge invariance and spontaneous symmetry breaking in the same theory, after a particular gauge transformation the Goldstone bosons disappear and the gauge bosons become massive. This mechanism [3,4] is usually called Anderson-Higgs mechanism which plays significant roles in both particle physics and condensed matter physics. One simple theory that exhibits the Anderson-Higgs mechanism is the Ginzburg-Landau phenomenological model of superconductivity [5]. When an Abelian gauge field is introduced the phase of the Ginzburg-Landau order parameter (the Goldstone boson) can be eliminated by an appropriate gauge transformation and the photon absorbs the phase and acquires a nonzero mass, which is interpreted as the Meissner effect in superconductivity. In performing this gauge transformation one crucial requirement is that the phase of the order parameter be an analytic function of space-time variables. This requirement is no doubt satisfied if we assume the phase is unique. However, this assumption may be too strict. Actually, the generally adopted assumption is that the order parameter is single-valued, which allows the phase either to be unique or to change by multiples of $2\pi $ after going around some closed path, depending on whether the superconductor is simply connected or not. In the former case we can always choose a gauge to reduce the phase to zero, but in the latter case it is impossible to do so, as we will show in the context. Therefore the conventional Anderson-Higgs mechanism does not apply to a multiply connected superconductor.

A superconductor may be multiply connected when it has geometrical holes in it or it, being a type-II superconductor, is in the Schubnikov state. Underdoped cuprates provide the third kind of multiply connected superconductors, in which the holes are caused by thermal phase fluctuations rather than mechanical means or strong external magnetic fields. As argued by Emery and Kivelson [6], underdoped cuprates, being doped Mott insulators with low superconducting carrier density and a relatively small phase stiffness, will undergo a Berezinskii-Kosterlitz-Thouless (BKT) [7,8] transition at some temperature $T_{{\rm BKT}}$, well below the temperature scale $T^{\ast }$ where the electron pairs are broken. At temperatures between $T_{{\rm BKT}}$ and $T^{\ast }$, free vortices appear as the topological excitations of phase fluctuations and destroy the long-range phase coherence among the pairs. In this region, called pseudogap region [9], the phase of the order parameter becomes a singular function and hence can not be eliminated by any gauge transformation.

In this Letter, we concentrate on the third kind of multiply connected superconductors and, in the pseudogap region, look for a substitute for the conventional Anderson-Higgs mechanism. It is known from the theory of BKT transition [10] that in the presence of free vortices the phase of the order parameter or the Goldstone boson in the Ginzburg-Landau model can be resolved into longitudinal and transverse components which are analytic and singular, respectively. Taking advantage of this fact, in this Letter we show that the longitudinal component can be eliminated by a special gauge transformation but the transverse component persists anyway. As a result, although the gauge field acquires a nonzero mass after absorbing the longitudinal component of the Goldstone boson, the equation for magnetic field does not exhibit Meissner effect. Thus, we obtain an extended Anderson-Higgs mechanism which exhibits completely new physics comparing with the old Anderson-Higgs mechanism.

We begin our discussion with the following Ginzburg-Landau model in two space and one time dimensions
\begin{equation}
{\cal L}=-\frac{1}{4}F_{\mu \nu }F^{\mu \nu }+\frac{1}{4m}D_{\mu }\phi
\left( D^{\mu }\phi \right) ^{\ast }-\alpha \left| \phi \right| ^{2}-\frac{%
\beta }{2}\left| \phi \right| ^{4},
\end{equation}
where the complex scalar field $\phi $ is the order parameter and $D_{\mu}\equiv \partial _{\mu }-2ieA_{\mu }$ is the covariant derivative. Here $m$ is the effective mass of a electron. The gauge field tensor $F_{\mu \nu }\equiv \partial _{\mu }A_{\nu }-\partial _{\nu }A_{\mu }$ as usual. For $\alpha >0$ the system stays in its symmetric phase and for $\alpha <0$ vacuum degeneracy and symmetry breaking take place, while the coefficient $\beta $ is always positive according to the Landau phase transition theory (note that it differs from BKT transition in most aspects).

The Lagrangian $\left( 1\right) $ is invariant under the local U(1) gauge transformation 
\begin{equation}
A_{\mu }\longrightarrow A_{\mu }-\partial _{\mu }\theta 
\end{equation}%
\begin{equation}
\phi \longrightarrow e^{2ie\theta }\phi ,
\end{equation}%
where $\theta \left( x\right) $ is an arbitrary coordinate-dependent function. 

In the symmetry breaking case, the ground state occurs at 
\begin{equation}
\left\langle \phi \right\rangle =\phi _{0}=\sqrt{-\frac{\alpha }{\beta }}%
\neq 0.
\end{equation}%
Expanding the Lagrangian (1) around this ground state, we write the scalar field as 
\begin{equation}
\phi \left( x\right) =\left( \phi _{0}+\eta \left( x\right) \right)
e^{i\theta \left( x\right) },
\end{equation}%
where $\eta \left( x\right) $ and $\theta \left( x\right) $ are the amplitude and phase fluctuations of the order parameter, respectively. Substituting the expression (5) into the Lagrangian (1), we get 
\begin{eqnarray}
{\cal L} &=&-\frac{1}{4}F_{\mu \nu }F^{\mu \nu }+\frac{1}{4m}\left( \partial
_{\mu }\eta \partial ^{\mu }\eta \right) +V\left( \phi _{0}+\eta \right) 
\nonumber \\
&&+\frac{1}{4m}\left( \phi _{0}+\eta \right) ^{2}\left( \partial _{\mu
}\theta -2eA_{\mu }\right) \left( \partial ^{\mu }\theta -2eA^{\mu }\right) ,
\end{eqnarray}
where $V(\phi )=$ $-\alpha \left| \phi \right| ^{2}-\frac{\beta }{2}\left|
\phi \right| ^{4}.$ The phase field $\theta \left( x\right) $, which enters the Lagrangian only through its gradient form, represents the massless Goldstone boson arising from symmetry breaking. Its appearance is a natural consequence of the spontaneous breakdown of a continuous symmetry [2] and seems to be a serious obstacle for the search of broken symmetry in nature because it has never been observed. However this obstacle may be overcomed if we introduce gauge invariance into the theory. Using the gauge invariance, we are able to adopt a special gauge (called the London gauge) to formally remove the phase $\theta \left( x\right) $ from the theory. In particular, letting 
\begin{equation}
A_{\mu }\longrightarrow A_{\mu }+\frac{1}{2e}\partial _{\mu }\theta ,
\end{equation}
we have 
\begin{eqnarray}
{\cal L} &=&-\frac{1}{4}\left( \partial _{\mu }A_{\nu }+\frac{1}{2e}\partial
_{\mu }\partial _{\nu }\theta -\partial _{\nu }A_{\mu }-\frac{1}{2e}\partial
_{\nu}\partial _{\mu }\theta \right)  \nonumber \\
&&\times \left( \partial ^{\mu }A^{\nu }+\frac{1}{2e}\partial ^{\mu
}\partial ^{\nu }\theta -\partial ^{\nu }A^{\mu }-\frac{1}{2e}\partial
^{\nu}\partial ^{\mu }\theta \right)  \nonumber \\
&&+\frac{1}{4m}\left( \partial _{\mu }\eta \partial ^{\mu }\eta \right)
+V\left( \phi _{0}+\eta \right) +\frac{e^{2}}{m}\left( \phi _{0}+\eta
\right) ^{2}A_{\mu }A^{\mu }.
\end{eqnarray}

For a simply connected superconductor, the phase $\theta \left( x\right)$ is an analytic function and it satisfies the equation
\begin{equation}
\partial_{\mu }\partial _{\nu }\theta -\partial _{\nu }\partial _{\mu }\theta \equiv 0,
\end{equation}
so the first term of (8) can be written as the original form $\frac{1}{4}F_{\mu \nu }F^{\mu \nu }$. Now we see that the massless Goldstone boson is removed completely by the gauge transformation (7) and the photon becomes massive. Note that the degrees of freedom of the theory does not decrease because the gauge field acquires a massive component which can be seen as another guise of the Goldstone boson. This is the conventional Anderson-Higgs mechanism which prevents the occurrence of the annoying massless Goldstone boson in an unexpected manner. Its generalization to non-Abelian groups provides a framework for the unified theory of the weak and electromagnetic interactions of the elementary particles.

On the other hand, for a multiply connected superconductor, in the vicinity of holes or vortices the superfluid velocity ${\bf v}_{s}=\frac{1}{2m}{\bf \nabla }\theta$ is not irrotational,
\begin{equation}
{\bf \nabla} \times {\bf \nabla} \theta \neq 0.
\end{equation}
It is obvious that in this case Eq.(9) is not satisfied. So, the first term of (8) can not be written as $\frac{1}{4}F_{\mu \nu }F^{\mu \nu }$. This is the reason why in a multiply connected superconductor it is impossible to choose a gauge where $\phi \left( x\right)$ is real. However in the pseudogap region of underdoped cuprates, the phase $\theta $ can be decomposed into two parts [10] 
\begin{equation}
\theta =\theta _{a}+\theta _{v},
\end{equation}
with $\theta _{a}$ the analytic spin wave (longitudinal) component and $\theta _{v}$ the singular vortex (transverse) component. The longitudinal and transverse components of $\theta $ have the properties [10]
\begin{equation}
{\bf \nabla} \times {\bf \nabla} \theta_{a}=0,
\end{equation}
\begin{equation}
{\bf \nabla} \times {\bf \nabla} \theta_{v} \neq 0.
\end{equation}
It is a well-known fact that the longitudinal and transverse components can be separated completely and there are no crossover terms in the action of the phase.

Now, using equation (11)-(13) and the fact that the longitudinal and transverse components of $\theta $ are independent of each other, we can make the following special gauge transformation 
\begin{equation}
A_{\mu }\longrightarrow A_{\mu }+\frac{1}{2e}\partial _{\mu }\theta _{a}.
\end{equation}
Since Eq.(12) tells us that 
$\partial_{\mu }\partial _{\nu }\theta_{a} -\partial _{\nu }\partial _{\mu }\theta_{a} \equiv 0$, we have 
\begin{eqnarray}
{\cal L} &=&-\frac{1}{4}F_{\mu \nu }F^{\mu \nu }+\frac{1}{4m}\left( \partial
_{\mu}\eta \partial ^{\mu }\eta \right) +V\left( \phi _{0}+\eta \right) 
\nonumber \\
&&+\frac{e^{2}}{m}\left( \phi _{0}+\eta \right) ^{2}\left[ A_{\mu }A^{\mu } +%
\frac{1}{4e^{2}}\partial _{\mu }\theta _{v}\partial ^{\mu }\theta _{v} -%
\frac{1}{e}A_{\mu }\partial ^{\mu }\theta _{v}\right] .
\end{eqnarray}
It can be seen from the Lagrangian that the longitudinal component $\theta_{a}$ is absorbed by the photon while the transverse component survives the gauge transformation (14). This may be understood as follows, although a massless photon can eat up the longitudinal component of $\theta $ to become massive, a massive photon can not eat anything more. Comparing with the case of simply connected superconductors, the Lagrangian (15) has one more degree of freedom corresponding to the topological excitation of the phase. Therefore, in addition to a mass term for gauge field, the theory also consists of a kinetic energy term for $\partial ^{\mu }\theta _{v}$ and an interaction term of gauge field and $\partial ^{\mu }\theta _{v}$. Thus, we obtain an extended Anderson-Higgs mechanism, which plays the same role in the pseudogap region of underdoped cuprates as the conventional Anderson-Higgs mechanism does in simply connected superconductors. The last two terms of (15) come from the singular vortex component of the phase and represent all unusual physics of the new mechanism.

That the longitudinal component of $\theta $ can be absorbed by the photon means that the longitudinal phase fluctuations are suppressed by electromagnetic fields, in both superconducting state and pseudogap state. This is in agreement with Millis and coworkers [11] who arrived at the same results based on field-theoretical perturbation techniques. On the other hand, since the transverse component of the phase still exists in the theory, it is reasonable to expect the transverse phase fluctuations be responsible for the unusual properties observed in pseudogap region [12].

Next we would like to derive the equation of the gauge field. For simplicity, we consider only a single vortex located at position {\bf x}=0 with vortex charge $q=1$. The generalization to the case of multiple vortices and higher charges is straightforward. From the Lagrangian (15) we derive the equations for ${\bf A}$ (neglecting the fluctuations of the amplitude $\eta $ around the ground state) as follows 
\begin{equation}
\nabla ^{2}{\bf A}-\mu {\bf A}=\zeta {\bf \nabla }\theta _{v},
\end{equation}%
where $\mu =-\alpha e^{2}/m\beta $ and $\zeta =\alpha e/m\beta $. In deriving the above equation, we have used the equality ${\bf \nabla } \cdot {\bf \nabla} {\theta_{v}}=0$. The equation for ${\bf A}$ without the inhomogeneous term is just the equation of the massive photon. In the case of a singly quantized vortex, the curl of the gradient of transverse component can be written as
\begin{equation}
{\bf \nabla \times \nabla }\theta _{v}=\frac{2\pi }{2m}{\bf {\hat {z}}}%
\delta \left( {\bf x}\right) ,
\end{equation}%
where ${\bf {\hat {z}}}$ is a unit vector perpendicular to the plane. Since
\begin{equation}
{\bf \nabla \times }\nabla ^{2}{\bf A=}\nabla ^{2}\left( {\bf \nabla \times A%
}\right) =\nabla ^{2}{\bf B}
\end{equation}
we obtain the equation for magnetic field%
\begin{equation}
\nabla ^{2}{\bf B-}\mu {\bf B=}\zeta \frac{2\pi }{2m}{\bf {\hat {z}}}\delta
\left( {\bf x}\right) .
\end{equation}
When the inhomogeneous term in the right-hand side of (19) is absent, we get the usual London equation which leads to the exponentially decaying behavior of the magnetic field in a superconductor and $\sqrt{1/\mu }=\sqrt{-\alpha e^{2}/m\beta }$ is the penetration depth. The inhomogeneous term on the right-side of hands changes significantly this behavior and prevents the occurrence of Meissner effect and superconductivity in the pseudogap region. Equation (19) tells us that in the place far from free vortices the Meissner effect still exists and magnetic field can not penetrate the material deeply. This is qualitatively agreement with the recent experiments [13,14].

It should be emphasized that, although calculations based on detailed dynamical models are needed in quantitatively understanding high-$T_{c}$ superconductors, we believe the extended Anderson-Higgs mechanism is more appropriate in explaining many material-independent properties including the suppression of longitudinal phase fluctuations by electromagnetic field at all temperatures, the dominance of transverse fluctuations in pseudogap region and the absence of Meissner effect in pseudogap region.

In conclusion, in this Letter we have obtained an extended Anderson-Higgs mechanism which works in the pseudogap region of underdoped cuprates. Firstly, in the presence of free vortices caused by strong thermal phase fluctuations, we show that the Goldstone field can not be eaten up by the photon and hence conventional Anderson-Higgs mechanism is not applicable. However, since the vortices are related to a BKT transition, we can resolve the Goldstone field into longitudinal and transverse components which are independent degrees of freedom. Then we show that the longitudinal component can be eliminated by a special gauge transformation while the transverse component of the phase stays in the theory owing to its singularity. As a result, the gauge field acquires not only a finite mass but also a direct coupling to the transverse superfluid velocity. We believe that this interaction can bring new physics [15] and hence deserves more detailed investigations. We also discuss the implication of the extended Anderson-Higgs mechanism and explain qualitatively several experimental facts observed in the pseudogap region. From the Lagrangian (15) we derive the equation for gauge field. It is different from that of the usual massive photon in that it has an inhomogeneous term, so in this region complete Meissner effect and superconductivity are not present.

Although till now we have been discussing the physics of superconductors, the applications of the Ginzburg-Landau model with U(1) gauge field are not restricted to superconductors. For example, in the context of hadron physics Nielsen and Olesen [16] suggested the Lagrangian (1) be a possible field theory that may contain a dual string structure, but they did not discuss the possibility of the decomposition (11) of the phase. It would be a very interesting task to find more physical systems where the decomposition (11) may be possible other than the underdoped cuprates discussed in this paper. Since vortices and transitions of BKT type are present in many systems, we believe the extended Anderson-Higgs mechanism is a general phenomenon in nature.

One of the authors (G.C.) is supported by the National Science Foundation in China.

\end{document}